\documentclass[12pt]{article}
\pdfoutput=1
\usepackage{amsmath,amssymb,amsfonts,epsfig,graphicx,euscript}%
\usepackage[pdftex,bookmarks,bookmarksnumbered,linktocpage,pdfstartview=FitH]{hyperref}
\hypersetup{colorlinks,%
citecolor=red,%
filecolor=blue,%
linkcolor=blue,%
urlcolor=blue,%
pdftex}
\usepackage{amsmath}
\usepackage{amsfonts}
\usepackage{graphicx}
\usepackage{caption}
\usepackage{subcaption}
\usepackage{float}
\usepackage[all]{hypcap}
\numberwithin{equation}{section}
\usepackage{cite}
\usepackage{calc}
\newlength{\spacer}
\newsavebox{\mybox}

\newcommand{\bse}{\begin{subequations}}
\newcommand{\ese}{\end{subequations}}
\newcommand{\be}{\begin{equation}}
\newcommand{\ee}{\end{equation}}
\newcommand{\bea}{\begin{eqnarray}}
\newcommand{\eea}{\end{eqnarray}}
\newcommand{\ba}{\begin{array}}
\newcommand{\ea}{\end{array}}
\newcommand{\h}{\frac{1}{2}}

\begin{document}
\begin{center}
{ \Large{\textbf{Electric Field Quench, Equilibration and Universal Behavior }}} 
\vspace*{1.5cm}
\begin{center}
{\bf  S. Amiri-Sharifi\footnote{s$_{-}$amirisharifi@sbu.ac.ir}, M. Ali-Akbari\footnote{m$_{-}$aliakbari@sbu.ac.ir}, H. R. Sepangi\footnote{hr-sepangi@sbu.ac.ir}}\\%
\vspace*{0.4cm}
{\it {Department of Physics, Shahid Beheshti University, G.C., Evin, Tehran 19839, Iran}}  \\
\vspace*{1cm}
\end{center}
\end{center}

\bigskip
\begin{center}
\textbf{Abstract}
\end{center}
We study electric field quench in ${\cal{N}}=2$ strongly coupled gauge theory, using the AdS/CFT correspondence. To do so, we consider the aforementioned system which is subjected to a time-dependent electric field indicating an out of equilibrium system. Defining the equilibration time $t_{eq}$, at which the system relaxes to its final equilibrium state after injecting the energy, we find that the rescaled equilibriation time $k^{-1}t_{eq}$ decreases as the transition time $k$ increases. Therefore, we expect that for sufficiently large transition time, $k\rightarrow\infty$, the relaxation of the system to its final equilibrium can be an adiabatic process. On the other hand, we observe a universal behavior for the fast quenches, $k\ll 1$, meaning that the rescaled equilibration time does not depend on the final value of the time-dependent electric field. Our calculations generalized to  systems in various dimensions also confirm universalization process which seems to be a typical feature of all strongly coupled gauge theories that admit a gravitational dual.

\newpage

\tableofcontents

\section{Introduction and Result}
Quantum quench is defined as the process of (rapidly) altering a physical coupling of a quantum system. Ever since the quantum quench became accessible in the lab experiments \cite{quench}, it has attracted a lot of attention to investigate its various properties. Another reason for its appeal is that, due to the time dependence of the coupling, the system is out of equilibrium and therefore understanding the evolution of the system is interesting enough. However, only with weak coupling assumption, analytic calculations have been done successfully \cite{q1}.

We are interested in considering a strongly coupled system in which a time-dependent electric field (rapidly) changes from zero to a finite value, \textit{electric field quench}. Since charged particles, such as quarks, exist in our system, applying a time-dependent electric field generates a current which non-trivially varies over time. The physical quantity we would like to study is \textit{equilibration time}. It describes the time at which the out of equilibrium system, where its characteristic is a non-trivial time-dependent current, approximately reaches its equilibrium specified by a constant current. In principle, this time depends on the final value of the electric field and on the transition time for which the electric filed reaches its final value. Because of the strongly coupled nature of the system, the perturbative methods are not applicable and thus the AdS/CFT correspondence (or more generally, the gauge-gravity duality) presents itself as a reasonable candidate.

According to the AdS/CFT correspondence type IIB string theory on  AdS$_5\times$S$^5$ background, resulting from near horizon limit of a stack of D3-branes,  corresponds to ${\cal{N}}=4$ super Yang-Milles (SYM) theory with the gauge group $SU(N)$ in $3+1$ dimensions \cite{Maldacena, Gubser:1998bc, Witten:1998qj, CasalderreySolana:2011us, Ramallo:2013bua}. In the large t'Hooft coupling constant and large $N$ limit, this correspondence reduces to a duality between a classical gravity (on asymptotically AdS$_5\times$S$^5$ backgrounds) and a strongly coupled SYM theory. In other words, the duality enables us to study various aspects of the SYM theories at strong couplings by using classical gravity. Applying this idea, quantum quenches in strongly coupled SYM theories with gravity duals have been extensively studied, for instance see \cite{Buchel:2013lla, Das:2011nk, Basu:2011ft, Buchel:2012gw, Buchel:2014gta, Astaneh:2014fga} and universal behaviour for fast quenches has been reported in \cite{Das:2014jna, Das:2014hqa, Caceres:2014pda, Buchel:2013gba}. As well, a generalization of the AdS/CFT correspondence states that a strongly coupled $SU(N)$ gauge theory in $p+1$ dimensions is dual to a classical gravity on the near horizon geometry of D$p$-branes \cite{Itzhaki:1998dd}.  

On the gauge theory side, in order to get a non-zero current we need charged particles, similar to quarks in the quantum chormodynamics, which transform according to the fundamental representation of  $SU(N)$ gauge group. From the view point of the AdS/CFT correspondence, on the gravity side the fundamental degrees of freedom can be added to the system by introducing $N_f$ D$q$-branes into the background, sourced by the D$p$-branes, in the probe limit by which we mean $N_f/N$ \cite{Karch:2002sh}. In fact, the probe limit guaranties that the back-reaction of the D$q$-branes on the background is negligible. Different aspects of these systems, i.e. D$p$-D$q$ systems, have been studied, see e.g. \cite{Myers:2006qr, Karch:2007pd}. The dictionary of the AdS/CFT correspondence states that the asymptotic behavior of the shape of the probe brane gives the mass of the fundamental matter and the expectation value of the corresponding operator in the gauge theory \cite{CasalderreySolana:2011us}. 

By choosing suitable configuration for the gauge fields living on the probe branes, one can consider a constant electric field on the brane and study its effect on the physical quantities such as the mass of the fundamental matter \cite{Nakamura:2010zd}. In an interesting paper \cite{Karch:2007pd}, it was shown that this electric field gives rise to a constant current on the brane and this current increases in the presence of $\alpha'$-correction \cite{AliAkbari:2010av}. Recently, the effect of time-dependent electric fields, which clearly indicate out of equilibrium system, has been studied in \cite{Hashimoto:2013mua} where it was shown that a time-dependent current is induced on the brane which finally relaxes to its equilibrium value, in accordance with  \cite{Karch:2007pd}.

As we already mentioned, we consider a ${\cal{N}}=2$ strongly coupled SYM theory which is subjected to a time-dependent electric field. The gravity dual is described by embedding a single ($N_f=1$) probe D7-brane in the AdS$_5\times$S$^5$ background when a time-dependent electric field is turned on the probe brane. We would like to study  the effect of the transition time $k$ and of the final value of the electric field $E_0$ on the equilibration time $t_{eq}$. Our results can be summarized as follows:
\begin{itemize}
\item Our calculations indicate that a universal behavior emerges when $k<1$, i.e. \textit{fast electric field quenches}. More precisely, by universality we mean the rescaled equilibration time $k^{-1}t_{eq}$ becomes independent of the final value of the electric field for small enough $k$.  

\item For $k>1$, i.e. \textit{slow electric field quench}, the rescaled equilibration time decreases as one raises $k$ at fixed final electric field value. In fact, it seems that for large enough values of $k$, the relaxation of the system to its final equilibrium is an adiabatic process. 
\end{itemize}
We then generalize the above configuration to the D$p$-D$q$ systems and discuss the effect of dimensions, in which the degrees of freedom live, on the equilibration time. We observe that both slow and fast electric field quenches qualitatively manifest a universal behavior for the systems we take into account. 

In Appendix \ref{general current}, we briefly review and explain the way that a \textit{time-independent} current can be found by using the gauge-gravity duality. In this paper we are working with the pulse function \eqref{kik2}, which doesn't have a continues second derivative, in Appendix \ref{newpulsefunctin} we introduce a new pulse function with a continues second derivative and discuss its physical outcomes. Our observations indicate that the main results remain unaffected.

\section{Time-Dependent Current}
Adding the fundamental matter (quark) to the ${\cal{N}}=4$ SYM is equivalent to consider the probe D7-brane in the asymptotically AdS background \cite{Karch:2002sh}. It is well-known that switching on a non-dynamical electric field on the probe D7-brane leads to a constant current at non-zero temperature as well as zero temperature \cite{Karch:2007pd}. The effect of a non-dynamical external electric field on the system, when the mass of the quarks has been supposed to be zero, has been studied in \cite{Hashimoto:2013mua} and it has been then generalized for massive quarks in \cite{Hashimoto:2014yza}. In the following a generalization of above system to D$p$-D$q$ systems will be considered.

To do so, we start with the following supergravity solution corresponding to near horizon limit of $N$ coincident D$p$-branes \cite{Itzhaki:1998dd}
\be\begin{split}\label{metric} %
 ds^2&=H^{-1/2}(-dt^2+dx_p{^2})+H^{1/2}(du^2+u^2 d \Omega_{8-p}^{2}),\cr
 d \Omega_{8-p}^{2}&=d\theta^2+\sin^2\theta d\Omega_{k}^2+\cos^2\theta d\Omega_{7-p-k}^2,
\end{split}\ee %
where 
\be %
 H(u)=(\frac{R}{u})^{7-p},\ \ \ e^{\phi}=H^{\frac{3-p}{4}},\ \ \ C_{01..p}=H^{-1}.
\ee 
The Dilaton field is represented by $\phi$ and $C_{01..p}$ is a $(p+1)$-form coupled to the D$p$-branes. The length scale $R$ can be written in terms of the string scale $l_s=\sqrt{\alpha'}$ and the string coupling constant $g_s=e^{\phi_{\infty}}$
\be %
 R^{7-p}=(4\pi)^{\frac{5-p}{2}}\Gamma(\frac{7-p}{2}) N g_s l_s^{7-p}.
\ee %
For the case $p=3$, the above solution reduces to $AdS_5\times S^5$. Since there is no decoupling limit for  $p \geq 5$, we consider only $p\leq 4$ cases. Apart form the case $p=3$, the Dilaton field is not constant, indicating that the gauge theory coupling constant is running and consequently the gauge theory does not enjoy  conformal symmetry \cite{Itzhaki:1998dd}.

According to gauge-gravity duality, a strongly coupled $SU(N)$ SYM theory living on the $(p+1)$-dimensional worldvolume of $N$ coincident D$p$-brane corresponds to a supergravity on the background \eqref{metric} in the large $N$ limit. The supergravity approximation is only reliable when the curvature and the string coupling constant are constrained to be small. These two conditions, in terms of a dimensionless coupling constant $g_{eff}$, result in
\be %
 1 \ll g_{eff} \ll N^{\frac{4}{7-p}},
\ee %
where %
\be\begin{split}
 g_{eff}&=g_{YM}^2 N u^{p-3}, \cr
 g_{YM}^2&=2\pi g_s(2\pi l_s)^{p-3}.
\end{split}\ee
We refer the reader to \cite{Itzhaki:1998dd} for more details.

The low energy effective action describing the dynamics of the probe D$q$-branes in an arbitrary background is given by 
\be\label{action}\begin{split}%
S&=S_{DBI}+S_{CS}, \cr
S_{DBI}&=-\tau_q \int d^8\xi e^{-\phi} \sqrt{-\det(g_{ab}+B_{ab}+(2\pi\alpha')F_{ab})}, \cr
S_{CS}&=\tau_q \int {\rm{P}}[ C^{(n)}e^B]e^{(2\pi\alpha')F}.
\end{split}\ee
Tension of the branes is $\tau^{-1}_q=(2\pi)^q g_s l_s^{q+1}$ and induced metric $g_{ab}$ and Kalb-Ramond $B_{ab}$ are 
\be\begin{split}%
g_{ab}&=G_{MN}\partial_a X^M \partial_b X^N, \cr
B_{ab}&=B_{MN}\partial_a X^M \partial_b X^N.
\end{split}\ee
$M, N, ..$($a, b, ..$) are used to describe the spacetime(worldvolume) coordinates. The background metric $G_{MN}$ is introduced in \eqref{metric} and $F_{ab}$ is the field strength of the gauge field living on the D$q$-branes. In the Chern-Simons part, ${\rm{P}}[..]$ is the pull-back of the bulk spacetime fields to the worldvolume of the probe brane and $C^{(n)}$ denotes the $(n+1)$-form Ramond-Ramond potential coupled to D$q$-branes. 

In order to compute a time-dependent current on the probe branes, it is more convenient to rewrite \eqref{metric} using the following change of coordinates 
\be\begin{split} 
 \rho&=u \sin\theta, \cr 
 \sigma&=u \cos\theta,
\end{split}\ee 
and we finally have 
\be
ds^2=H^{-\h}(-dt^2+dx_p^2)+H^\h\Big(d\rho^2+\rho^2d\Omega_k^2+d\sigma^2+\sigma^2d\Omega^2_{7-p-k}\Big).
\ee
Now let us consider the following embedding for the D$p$-D$q$ system 
\be\label{config}
\begin{array}{cccccccccccc}
& t & x_1 & .. & x_d & x_{d+1} & .. & x_p & \rho & \Omega_k & \sigma & \Omega_{7-p-k}  \\
Dp & \times & \times & \times & \times & \times & \times &  \times &  &   &   &  \\
Dq & \times & \times & \times & \times &  &  &  & \times & \times &  & ,
\end{array}
\ee 
where $q=k+d+1$. Now we need to introduce a suitable ansatz to study a time-dependent current on the brane. Frist we consider the \textit{massless case} meaning that all the transverse coordinates must be zero. In the presence of an external time-dependent electric field on the brane, we expect a time-dependent current $\langle J^x \rangle$. According to AdS/CFT dictionary, the current $J^x$ is dual to the $U(1)$ gauge field on the brane. Therefore, in our ansatz, we include
\be %
 A_x(t,z)=-\int^{t} E(t') dt'+a(t,z),
\ee %
where a new radial coordinate $z=\frac{2}{5-p} R^{-\frac{p-7}{2}} u^{\frac{p-5}{2}}$ is applied. Near the boundary  $z\rightarrow 0$, the gauge field asymptotically approaches \cite{Karch:2007pd}
\be\label{current0} %
 a(t,z)=a_0(t)+\frac{\langle J^x(t) \rangle}{2{\cal{N}}(2\pi\alpha')^2}z^2+O(z^4),
\ee %
where ${\cal{N}}=\tau_q (2\pi)^2$ \cite{Karch:2007pd} and it is then easy to see that 
\be\label{current} %
 \langle J^x(t) \rangle \propto \lim_{z\rightarrow 0} \partial_z^2 a(t,z).
\ee %

In the background we are interested in, the Kalb-Ramond field is zero and regarding \eqref{config} configuration, the form field in the background does not couple to the D$q$-brane. Therefore, the CS-part of the action and the second term under the square root in \eqref{action} do not contribute to the action describing the dynamics of the probe branes. Moreover, in order to have a stable background at zero temperature we
consider supersymmetric systems where $q = p + 4, p + 2, p$ and $k = 3, 2, 1$,
respectively. For such configurations we have $p-q+2(k-1)=0$. With the above assumptions the DBI action for the background \eqref{metric} reduces to the following Lagrangian
\be\label{lagranigian}\begin{split}
{\cal{L}} & \propto z^\beta \sqrt{1- \gamma(2\pi\alpha')^2 (\frac{z}{R})^\alpha[(\partial_t A_x)^2-(\partial_z A_x)^2]},\cr
&= z^\beta \sqrt{w},
\end{split}\ee
where $S_{DBI} = \int d^8\xi {\cal{L}}$ and %
\be\begin{split}
\alpha= \frac{2(p-7)}{p-5}\, ,\ \ \beta = \frac{q-2p+9}{p-5}\, ,\ \ \gamma= (\frac{p-5}{2})^{\alpha}.
\end{split}\ee
A few values of $\alpha, \beta$ and $\gamma$, depending on the choice of $(p,q)$ system, has been listed in Table 1.
\begin{table}[ht]
\caption{Parameters}
\centering
\begin{tabular}{c c c c c c}
\hline\hline 
Case & $p$ & $q$ & $\alpha$ & $\beta$ & $\gamma$ \\[0.5ex]
\hline
1 & 3 & 7 & 4 & -5 & $1$ \\
2 & 3 & 5 & 4 & -4 & $1$ \\
3 & 4 & 6 & 6 & -7 & $1/64$\\ 
4 & 2 & 4 & $10/3$ & -3 & $(3/2)^{10/3}$ \\[1ex]
\\
\hline
\end{tabular}
\end{table}

The equation of motion for $A_x$ derived from the Lagrangian \eqref{lagranigian} is %
\be\label{eom}
 \partial_z\left(\frac{z^{\alpha + \beta}\partial_z A_x}{\sqrt{w}}\right)-\partial_t\left(\frac{z^{\alpha +    \beta}\partial_t A_x}{\sqrt{w}}\right)=0.
\ee
Our aim is to solve the above equation of motion for $A_x$ for arbitrary values of $p$ and $q$ using the numerical methods. To do so. we need to specify the time dependence of the external electric field. We choose two functions 
\bse\begin{align}
 \label{kik1} E(t) &=\frac{E_0}{2}\left(1+\tanh(\frac{t}{k})\right), \\
 \label{kik2}\qquad E(t) &= E_0
  \begin{cases}
   0 & \text{if } t \leq 0,\\
   \frac{1}{2}\left(1-\cos({\frac{\pi t}{k}})\right) & \text{if } 0 \leq t \leq k, \\
   1       & \text{if } t \geq k,
  \end{cases}
\end{align}\ese %
and will discuss their similarities and differences in the next two sections. The first function shows that the electric field is zero at $t=-\infty$ and it reaches $E_0$ at $t=+\infty$. However, \eqref{kik2} is zero from $-\infty$ to zero. At $t=0$ it starts changing to reach a maximum value at $t=k$ and remains constant afterwards. This function provide a more realistic alternative for \eqref{kik1}, as its behavior gives a more convenient meaning to the concept of \textit{turning on} an electric field and, after certain "finite" amount of time, reaching a maximum value $E_0$. The function \eqref{kik2} has been used to describe a time-depend meson melting in the AdS-Vaidya background in \cite{Ishii:2014paa} and its generalization in the presence of an external constant magnetic field has been done in \cite{Ali-Akbari:2015bha}. Notice that in both functions, $k$ is a measure to indicate how fast the functions can reach a maximum value. In other words, $k$ represents the transition time from zero to finite value. For instance, a small (large) value of $k$ indicates a fast (slow) growth in the electric field.

\section{D3-D7 Case}\label{37system}

\begin{figure}
  \includegraphics[width=65mm]{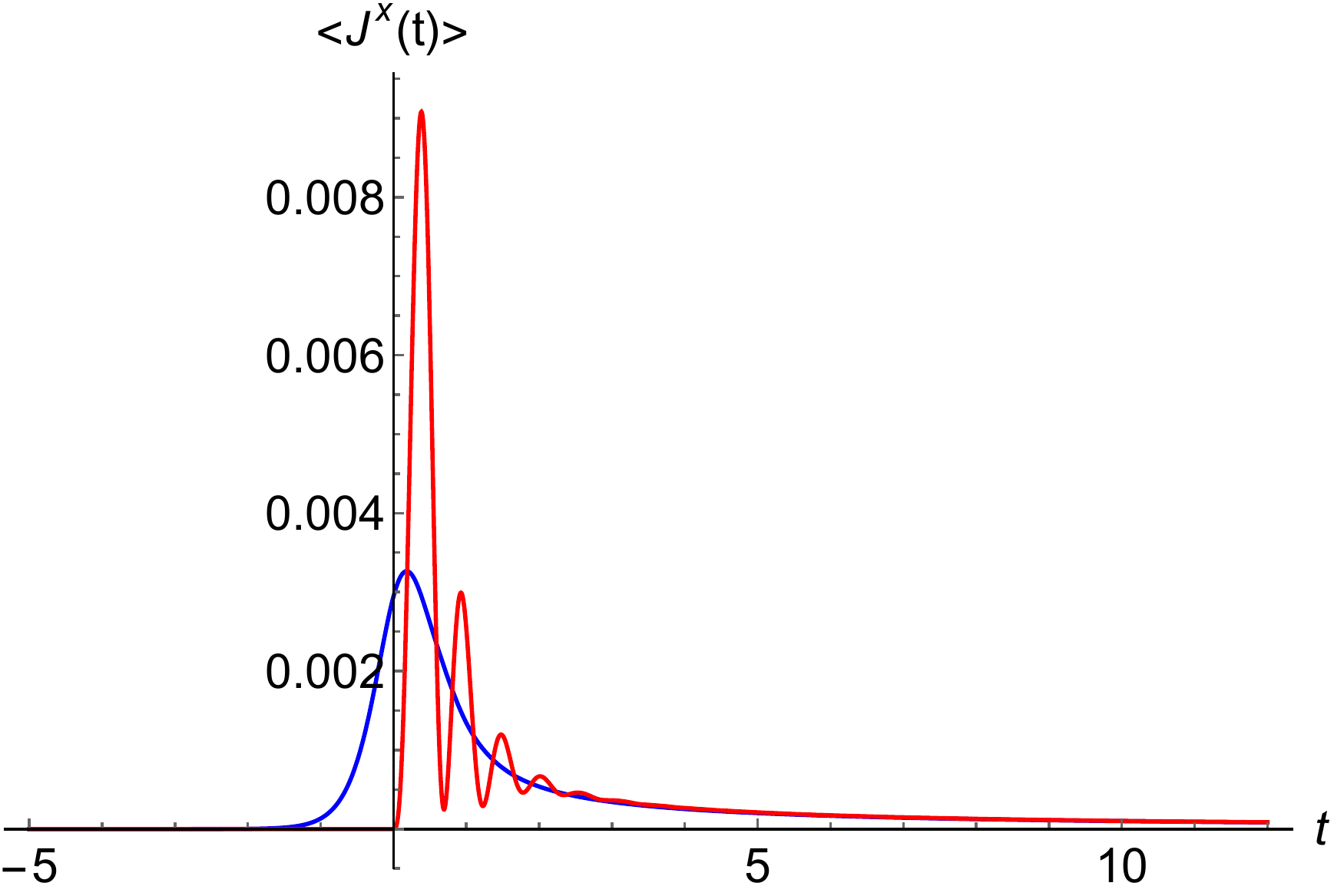}
  \hspace{2mm}
  \includegraphics[width=65mm]{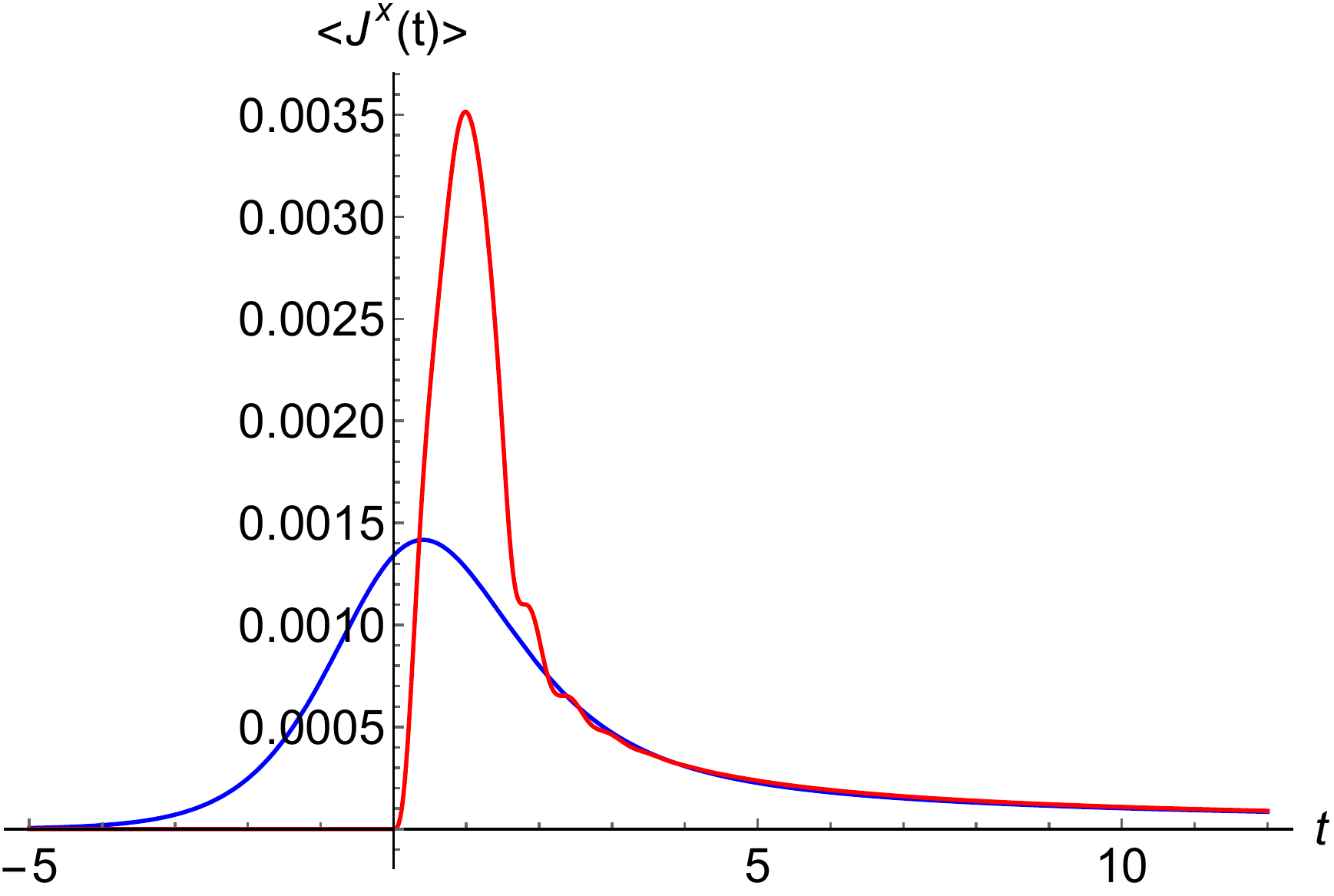}
  \includegraphics[width=65mm]{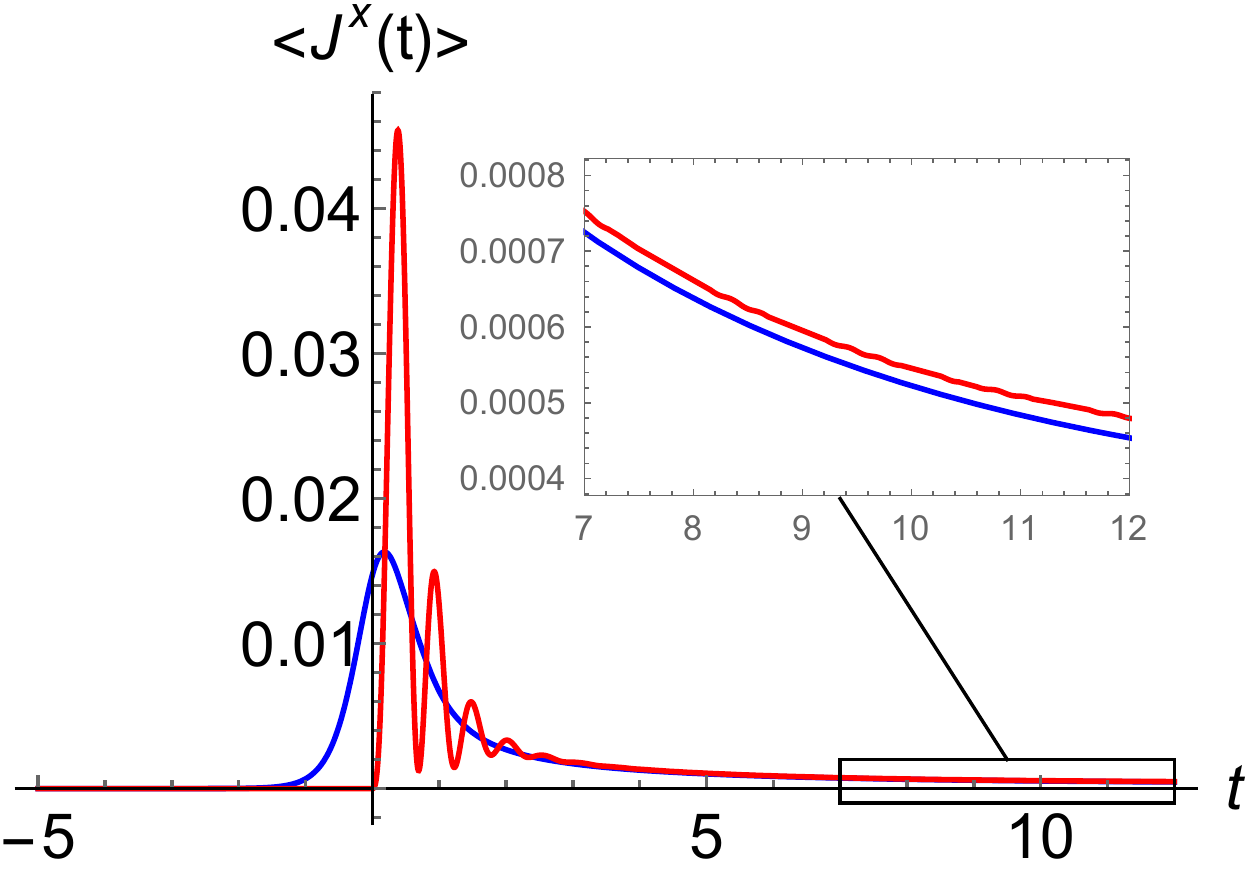}
  \hspace{2mm}
  \includegraphics[width=65mm]{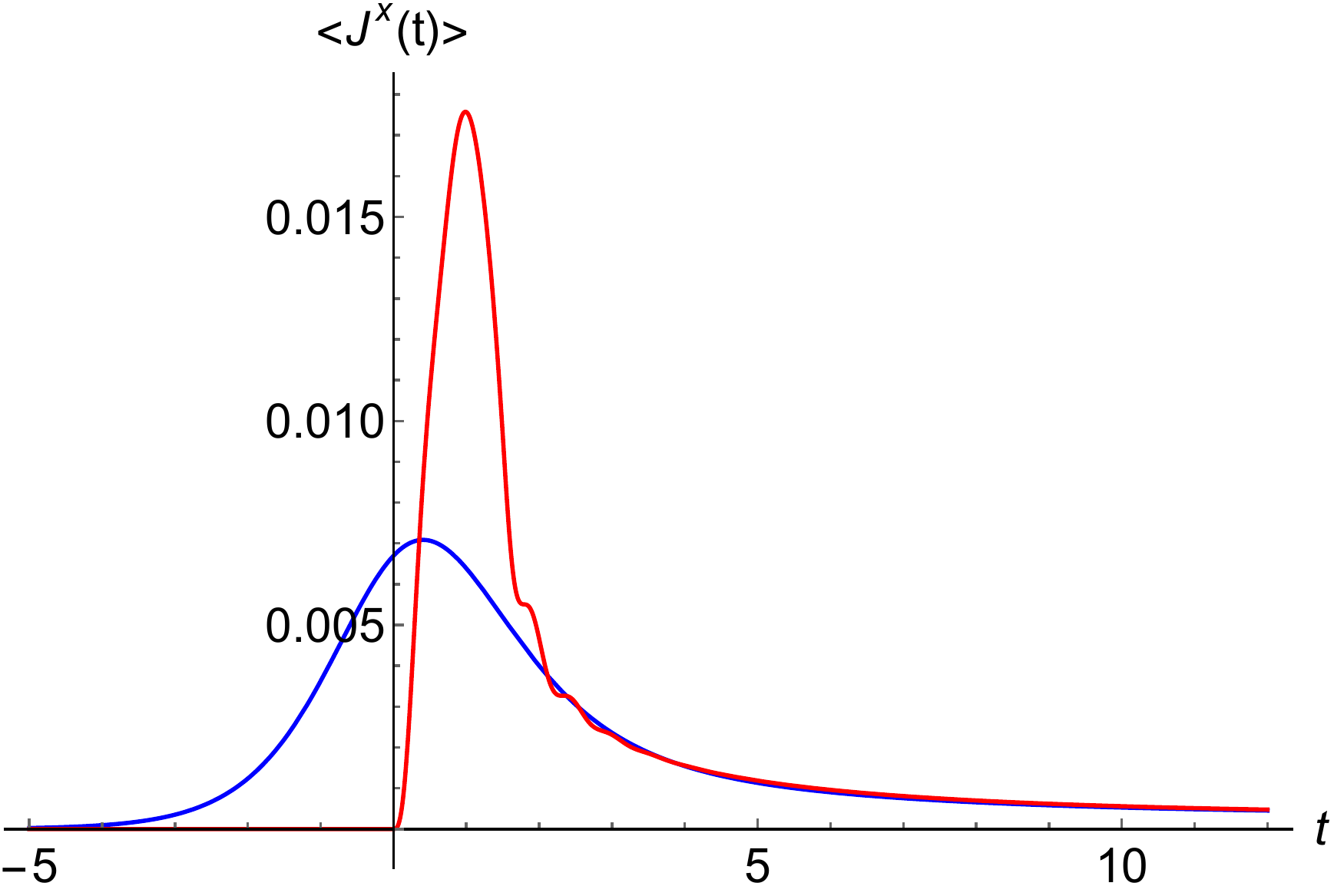}
\caption{Numerical results for current as a function of time for different values of electric fields, transition times and pulse functions.\\
top: $E_0=0.001$ with $k=0.5$ (left) or $k=1.5$ (right), \\ 
bottom: $E_0=0.005$ with $k=0.5$ (left) or $k=1.5$ (right). }\label{5Jt}
\end{figure}
\begin{figure}
\begin{center}
 \includegraphics[width=65mm]{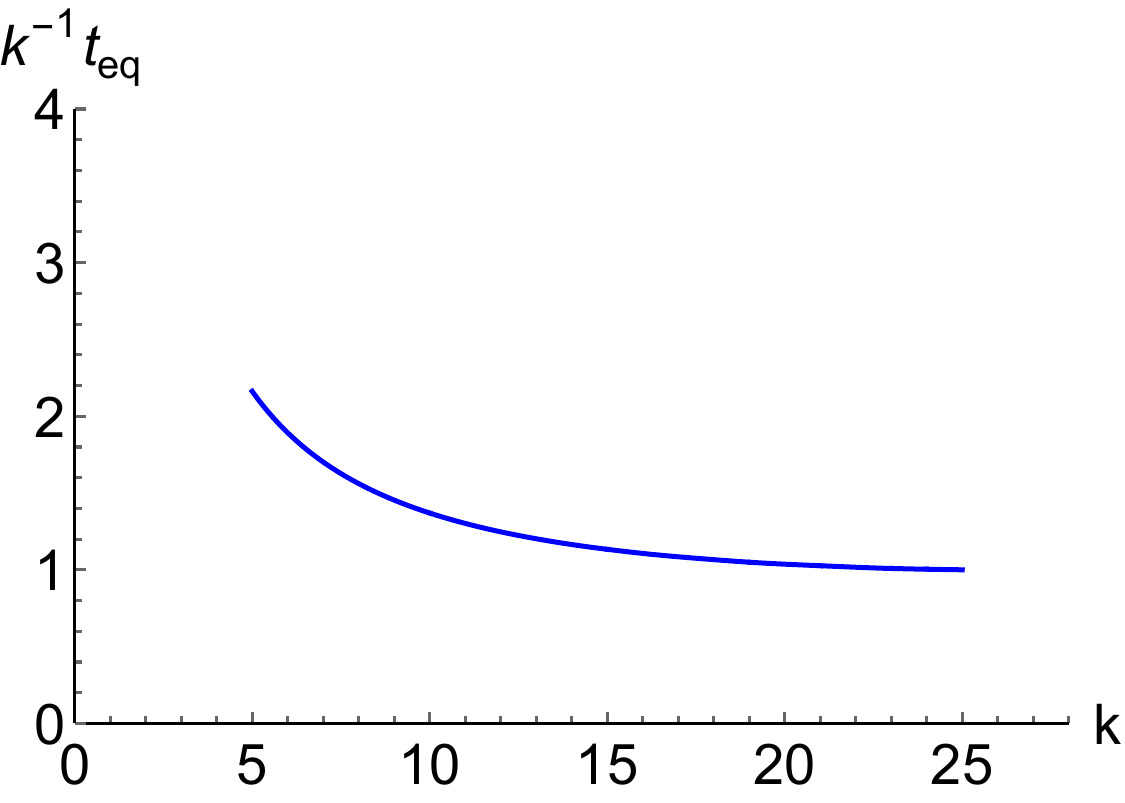}
 \hspace{2mm}
 \includegraphics[width=65mm]{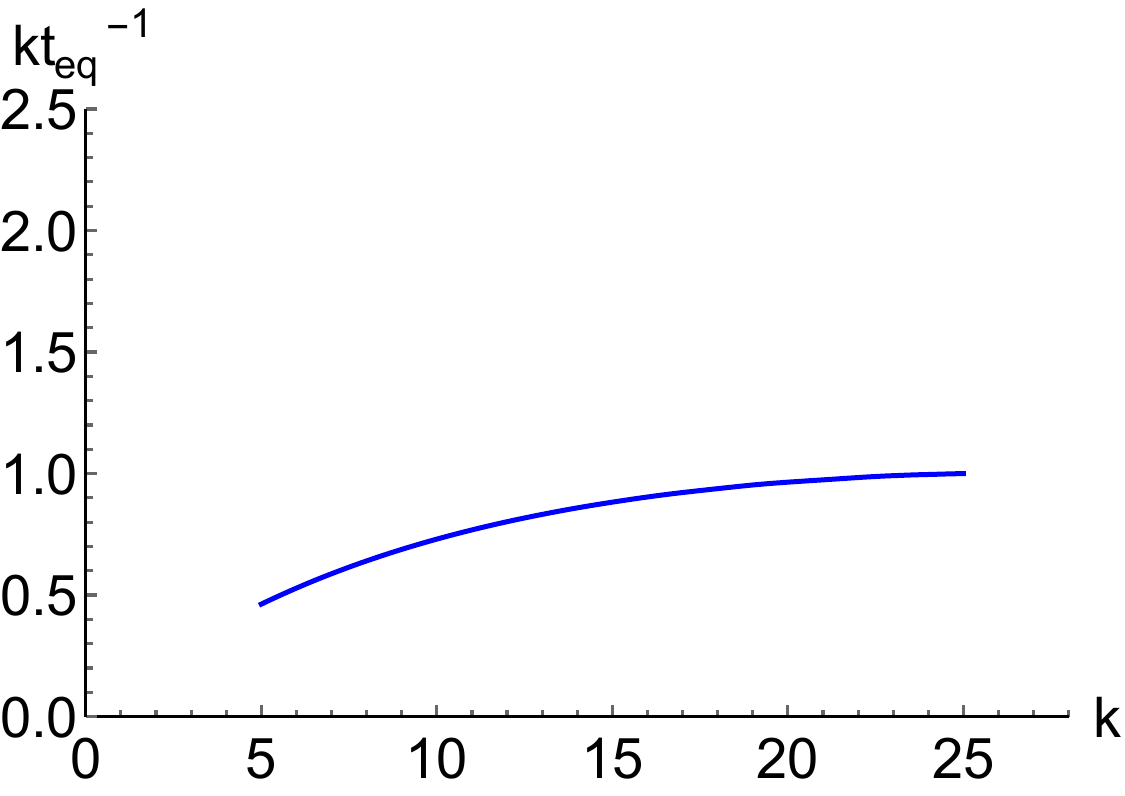}
 \caption{The left panel shows the resacled equilibration time versus the transition time. The right panel indicates that $(k^{-1}t_{eq})^{-1}$ goes to zero for smaller values of transition times, as we shall see for fast quenches. }\label{thermalization}
\end{center}
\end{figure}

Properties of the $(3,7)$ system which is subjected to the time-dependent function \eqref{kik1} were studied in \cite{Hashimoto:2013mua}. According to \eqref{current0} and \eqref{current}, due to a time-dependent electric field, a time-dependent current is induced on the brane. At early times the system is highly out of equilibrium and the current reaches its equilibrium value asymptotically. The equilibration time can be defined as the time at which the final value of the time-dependent current is equivalent to its equilibrated value. In order to obtain the equilibration time, as it was mentioned in the previous section, we will use the pulse function \eqref{kik2}. Since this function does not have a continuous second derivative, in Appendix \ref{newpulsefunctin} we introduce another pulse function which has a continuous second derivative and study its features.

As it was expected and is clearly seen from figure \ref{5Jt}, the amplitude of the pulse functions \eqref{kik1} and \eqref{kik2}, $E_0$, and the time of energy injection, $k$, control the large peak of the time-dependent current, for instance, at fixed $k$ by increasing $E_0$ the initial current peak rises. Similarly number of oscillations increases when one considers higher $E_0$. On the contrary, as one increases $k$, the value of the initial large peak and the number of oscillations decrease for a fixed value of $E_0$. Moreover, our investigations show that the independence of the final current from the choice of the pulse functions, is a universal feature of the system, as it was expectable. On the other hand, different behavior of the system at early time, due to different pulse functions, indicates that the system is sensitive to the manner of energy injection. As expected, different methods of injecting energy to the system would cause different reactions during this early time interval. 

We continue this section by studying the equilibration time. In \cite{Hashimoto:2013mua}, it was discussed that the appearance of an apparent horizon on the D7-brane  signals the thermalization on the brane. In fact the earliest time that an observer sitting at the boundary can see the apparent horizon formation in the bulk is considered as the time-scale of the thermalization.  More explicitly, a time-scale for the thermalization in terms of the location of the apparent horizon has been defined which the authors claim that it behaves universally; meaning that the time-scale of the thermalization is independent of $k$. 

Here, following  \cite{Buchel:2012gw, Buchel:2013lla, Buchel:2014gta}, we define  %
\be %
 \epsilon(t)\equiv\bigg\lvert\frac{\langle J^x_{eq}\rangle-\langle J^x(t)\rangle}{\langle J_{eq}^x\rangle}\bigg\rvert,
\ee %
where $\epsilon(t)$ denotes the relative error for the current. $J^x_{eq}$ is the final equilibrated current that $J^x(t)$ relaxes to in the long time limit. Therefore the equilibration time can be considered as the time that $\langle J^x(t)\rangle$ deviates less than $5\%$ from its final equilibrium value, i.e. $\epsilon({t_{eq}})<0.05$ and stays blew this limit thereafter. Notice that $\langle J^x_{eq}\rangle$ is given by $R \langle J^x_{eq}\rangle=(2\pi\alpha' E)^{\frac{3}{2}}$ \cite{Karch:2007pd} for D3-D7 system (see also Appendix \ref{general current} for more details). We found that the response of the system to the energy injection becomes different for two cases, $k>1$ and $k<1$. We will summarize the results in the following.

%
%
%
\begin{itemize}
\item{\textbf{Slow Quench}}\\
In the case of $k>1$, our numerical results are shown in figure \ref{thermalization}. It is clear that by increasing the value of the transition time $k$, the rescaled equilibration time $k^{-1}t_{eq}$ decreases. In fact, when for example $k\sim 25$, the equilibration time becomes comparable to $k$ and therefore the system doesn't have enough time to equilibrate during the transition. For smaller values of $k$, the system needs more time to equilibrate after finishing the injection of energy.  

A comparison between a time-dependent current and an equilibrated current, which in our unit is $\langle J^x_{eq}\rangle=E^{3/2}$, has been plotted in figure \ref{slow}(left). One can see that $\delta J= |\langle J^x(t) \rangle - \langle J^x_{eq}\rangle|$, which is a measure of deviation from equilibrium, is noticeable for small $k$.  However, for larger values of the transition time this deviation becomes smaller and hence one expects that $\delta J \rightarrow 0$ when the transition time goes to infinity. Therefore, It is expected that for sufficiently large value of $k$, \textit{i.e.} $k \rightarrow \infty$, the equilibration time becomes much smaller than the transition time indicating that the degrees of freedom living on the brane have enough time to thermalize during the transition. As a result, in this limit the system passes through approximately equilibrated situations and consequently the above limit resembles an adiabatic limit. In the right picture the maximum value for $\delta J$ is depicted versus $k$, which clearly shows the smooth flattening of the graphs in the left figure.

\begin{figure}
\begin{center}
  \includegraphics[width=65mm]{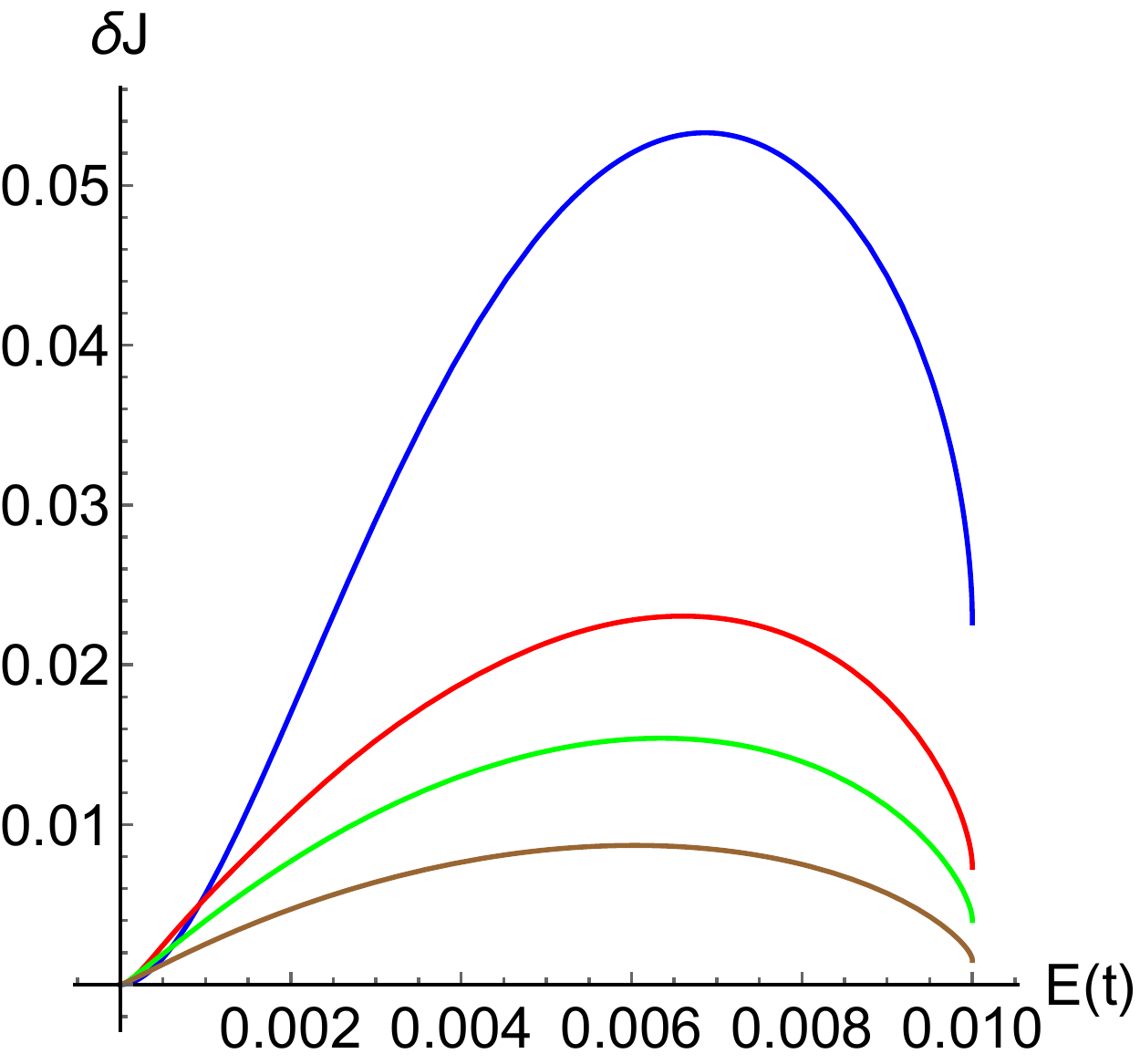} \quad
  \includegraphics[width=65mm]{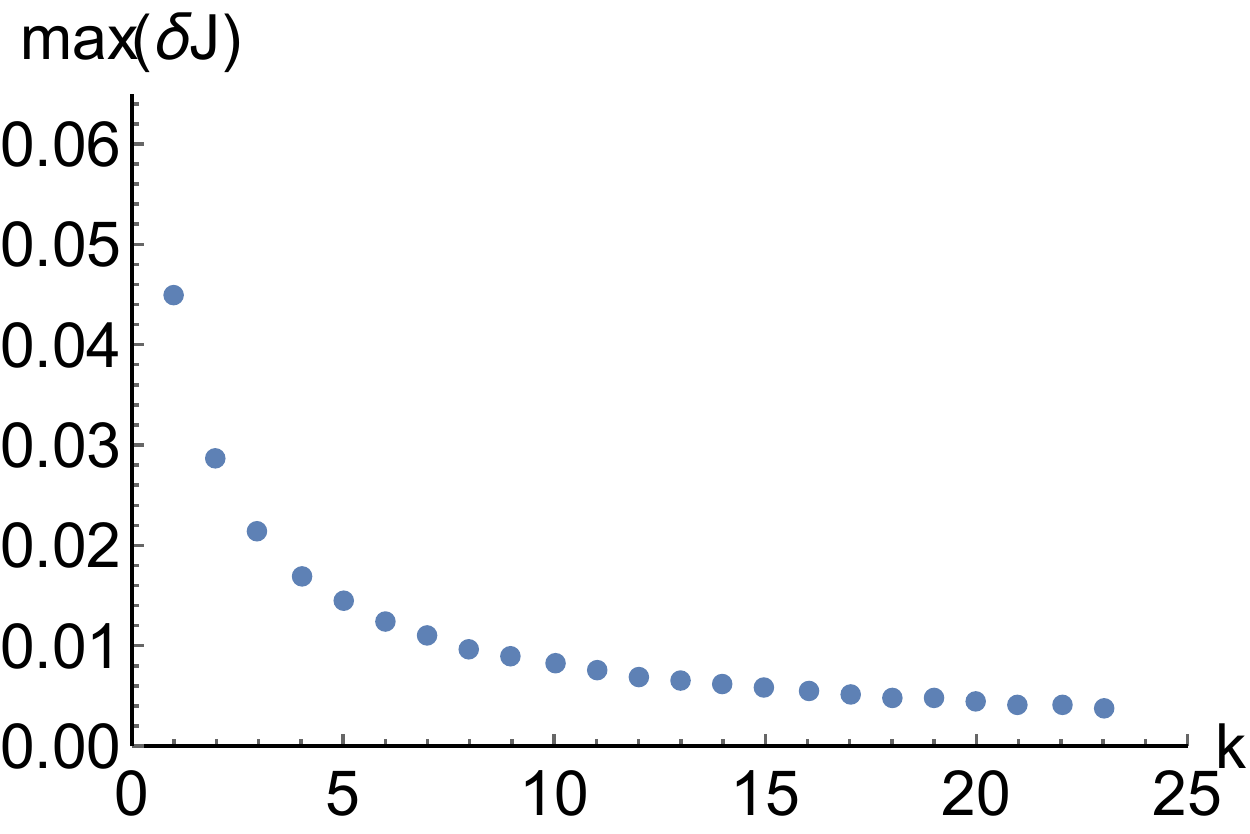}
\caption{Left: $\delta J= |\langle J^x(t) \rangle - \langle J^x_{eq}\rangle|$ in terms of $E(t)$ has been plotted for $k=1$ (blue), $k=3$ (red), $k=5$ (green) and $k=10$ (brown) (top to bottom). Right: the maximum of $\delta J$ versus $k$.}\label{slow}
\end{center}
\end{figure}

\begin{figure}
\begin{center}
  \includegraphics[width=65mm]{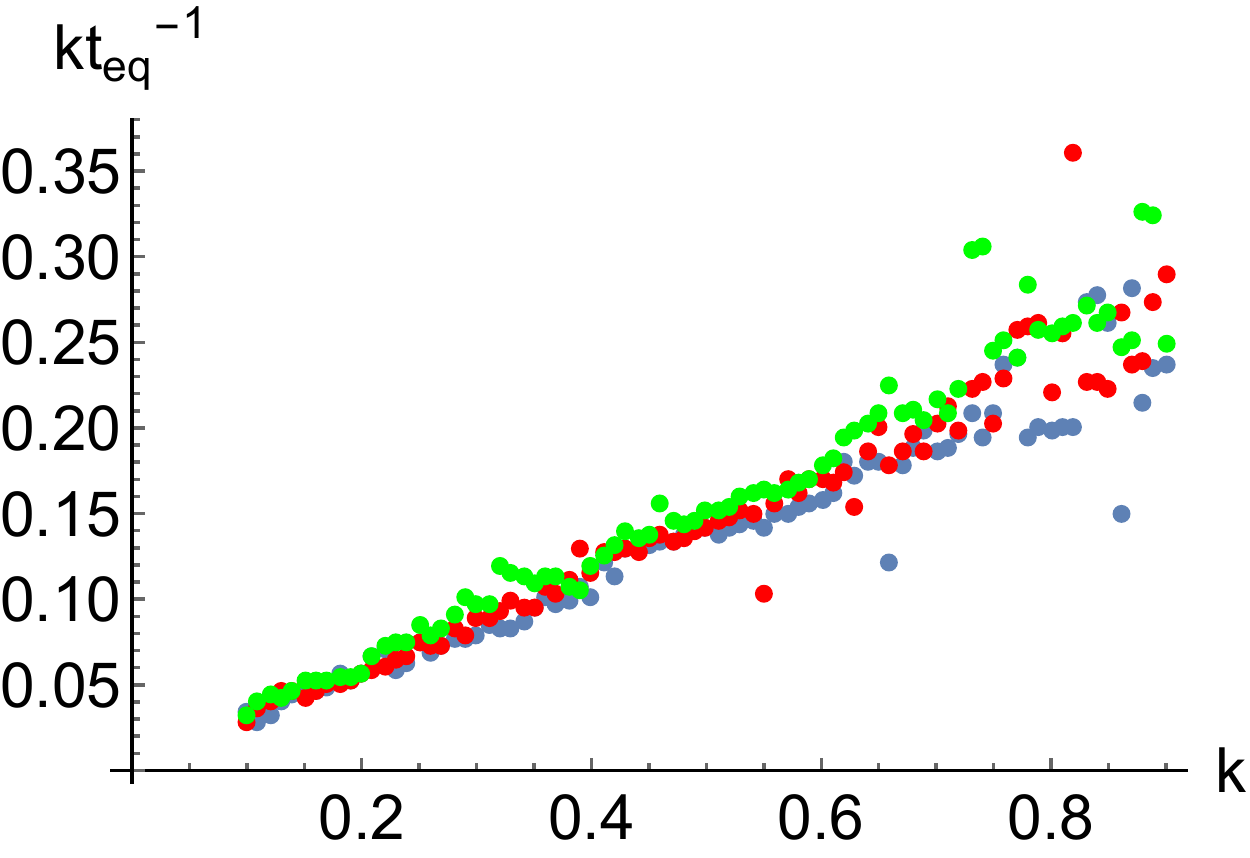}  \quad
   \includegraphics[width=65mm]{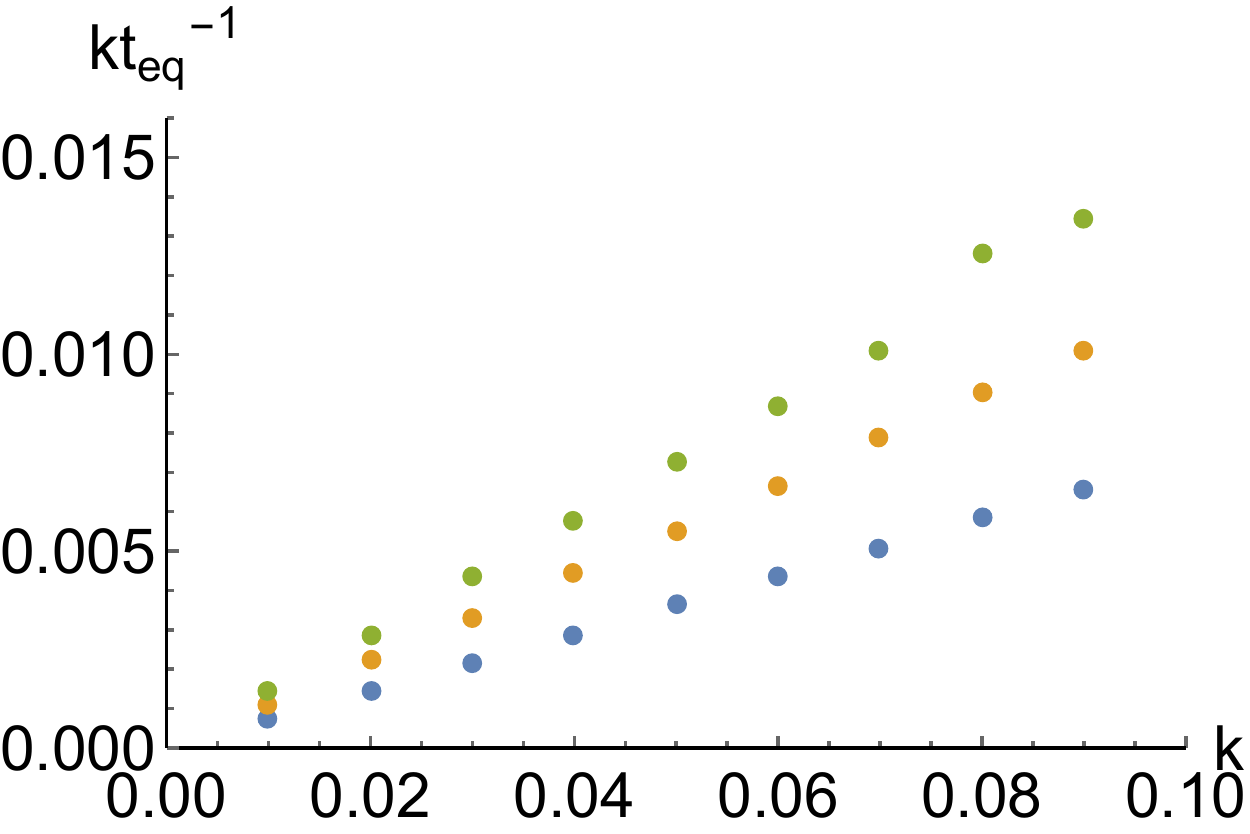}
\caption{The inverse of the resacled equilibration time versus the transition time for $E_0=0.24$ (blue), $E_0=0.27$ (red) and $E_0=0.31$ (green). }\label{fast}
\end{center}
\end{figure}


\item{\textbf{Fast Quench}}\\
The behavior of the equilibration time in the region with $k<1$ is interesting enough to investigate separately. This behavior can be seen in the data shown in figure \ref{fast}. As seen in these figures, when $k\lesssim 1$, for three different values of the electric field, their equilibration time is not the same. By decreasing the transition time, one can see that the difference among equilibration times become smaller.  For small enough values of $k$, the figure shows that the equilibration time for these three values of the electric field become the same or in other words it seems that the equilibration time is \textit{independent} of the final values of the electric field in the case of fast quench and it is set by the transition time $k$, only. 
\end{itemize}

Gravitational collapse of a massive scalar field in a time-dependent background which is asymptotically AdS spacetime has been studied in \cite{Buchel:2014gta}. In the field theory side, the mentioned configuration corresponds to a thermal quench in a strongly coupled ${{\cal{N}}=4}$ SYM theory. Using two-point correlation functions of suitable operators and the evolution of the entanglement entropy of the above system, the equilibration time has been computed. Interestingly, similar to our case, for slow quenches the rescaled equilibration time decreases with an increase in  the transition time. In the case of fast quenches, a universal behavior is also reported in the literature, for instance see \cite{Buchel:2014gta, Das:2014jna, Das:2014hqa, Caceres:2014pda, Buchel:2013gba}.

We also like to emphasize on a significant point here. In \cite{Buchel:2014gta} one should solve the equations of motion for the background metric and scalar field, simultaneously. In other words, the back-reaction of the scalar filed has been considered on the background and therefore the Einstein's equations describe dynamics of the system. On the other hand, in our case we are working in the probe limit and we ignore the back-reaction of the probe brane on the background. Therefore, DBI action describes dynamics of the system in our case. Despite the fact that the actions and equations of motion are different, the slow and fast quenches of the scalar field and electric field behave similarly. One may conclude that the universality observed in these results happen is generic property of strongly coupled theories with gravity duals, similar to the universality of shear viscosity \cite{CasalderreySolana:2011us}.  


\section{Generalization to D$p$-D$q$}
Now one can question whether the dimensions, in which the fundamental matter or SYM theory degrees of freedom live, affect the final physical results such as equilibration time and the universal behavior at small $k$. In order to find out the answer, we consider D$p$-D$q$ systems where depending on the embeddings of each brane, the various degrees of freedom can live on different dimensions. According to AdS/CFT dictionary, in the case of D3-D7 system the SYM theory degrees of freedom, in the adjoint representation of the corresponding gauge group, as well as fundamental matter are living on the 3+1 dimensions. A generalization of AdS/CFT dictionary to D$p$-D$q$ system has been extensively studied, for example see \cite{Karch:2007pd, Myers:2006qr}. In short, for the supersymmetric cases, $q=p+4$, $q=p+2$ and $q=p$, the SYM theory degrees of freedom live on $p+1$ dimensions. However, the fundamental matter is confined to a $d+1$-dimensional defect, see \eqref{config}, where D$p$- and D$q$-branes coincide. In the following we consider three interesting cases.

\begin{figure}
\begin{center}
  \includegraphics[width=65mm]{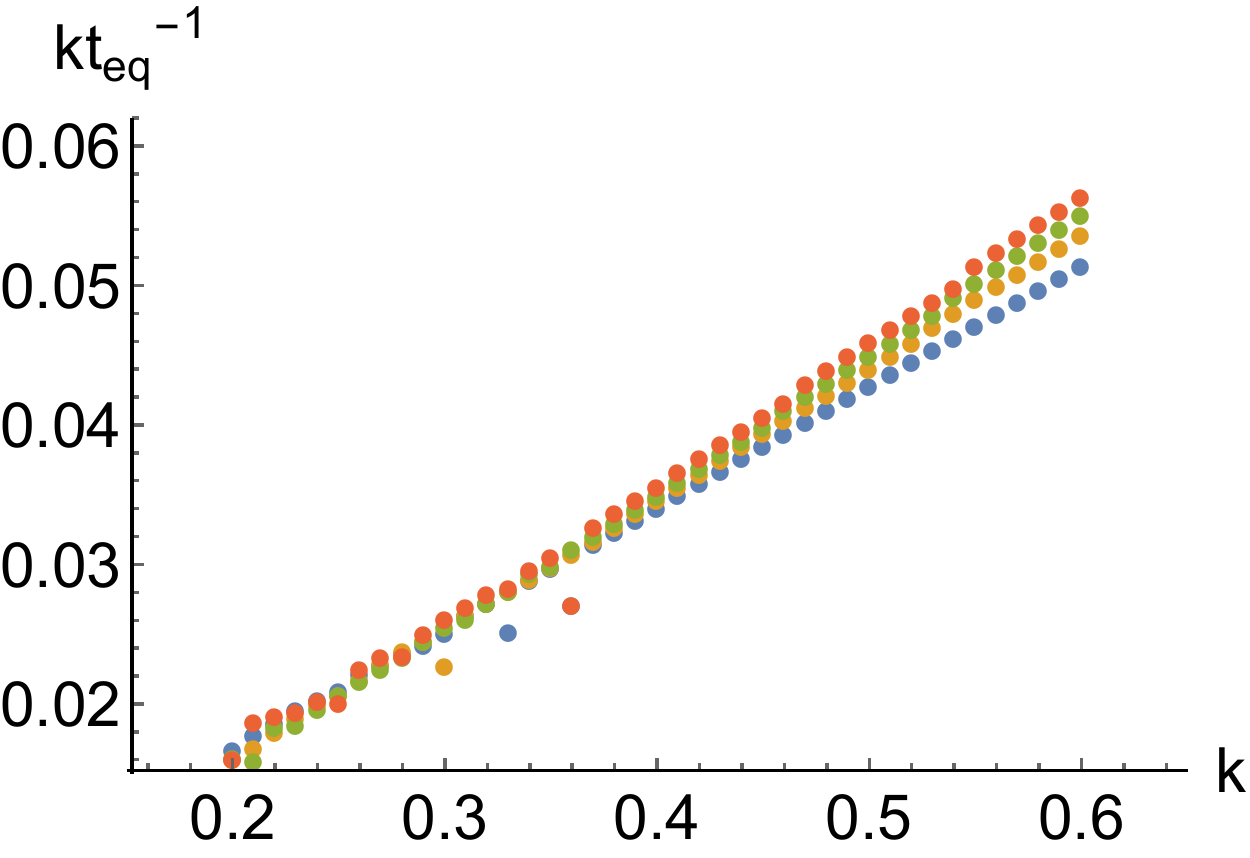}
  \hspace{4mm}
  \includegraphics[width=65mm]{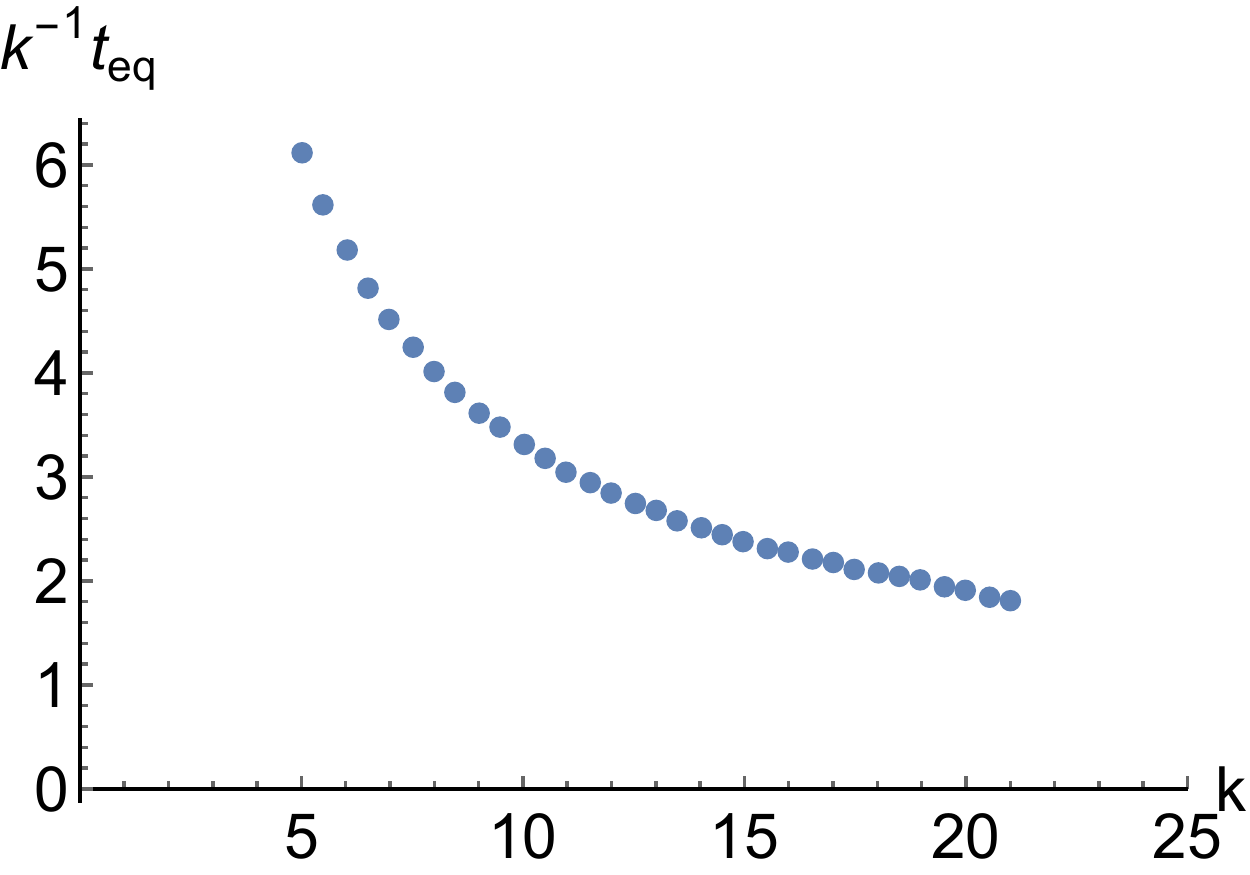}  
  \includegraphics[width=65mm]{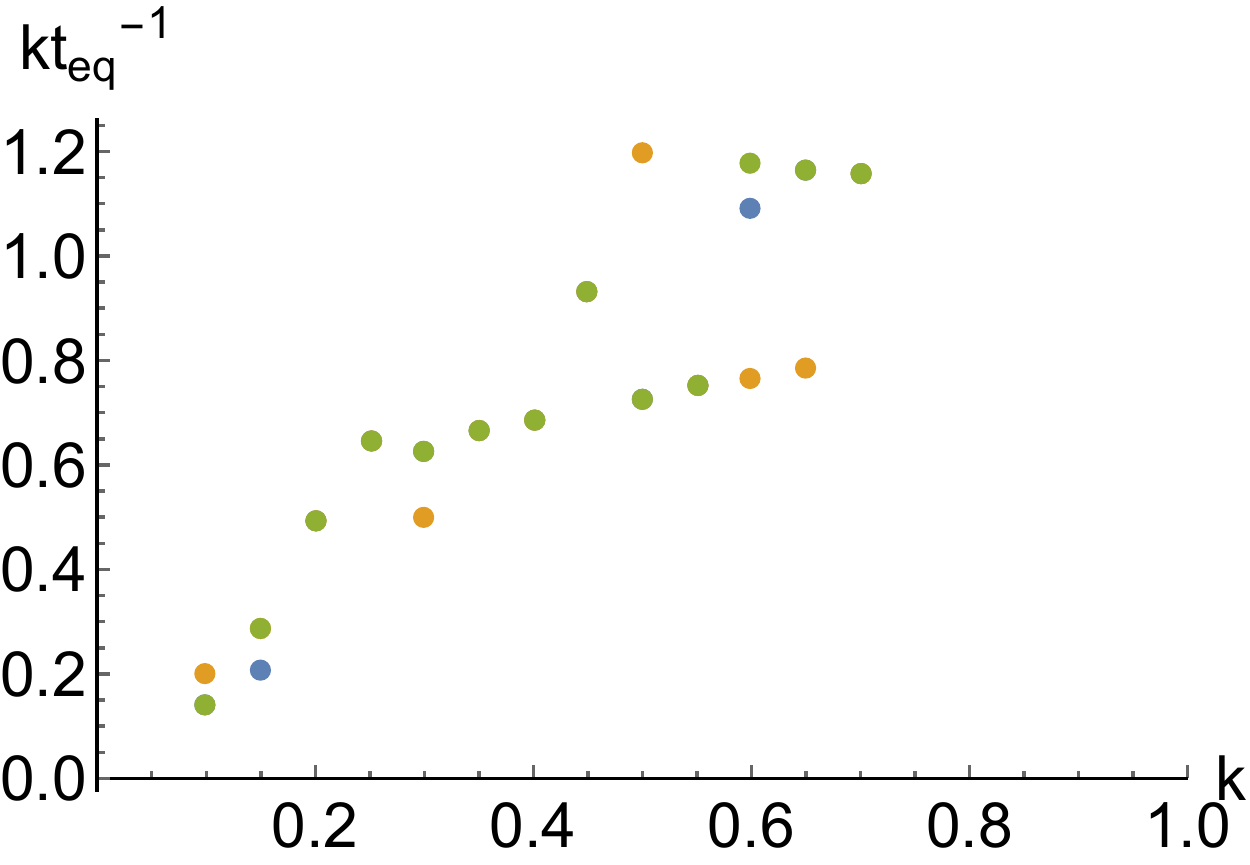}
  \hspace{4mm}
  \includegraphics[width=65mm]{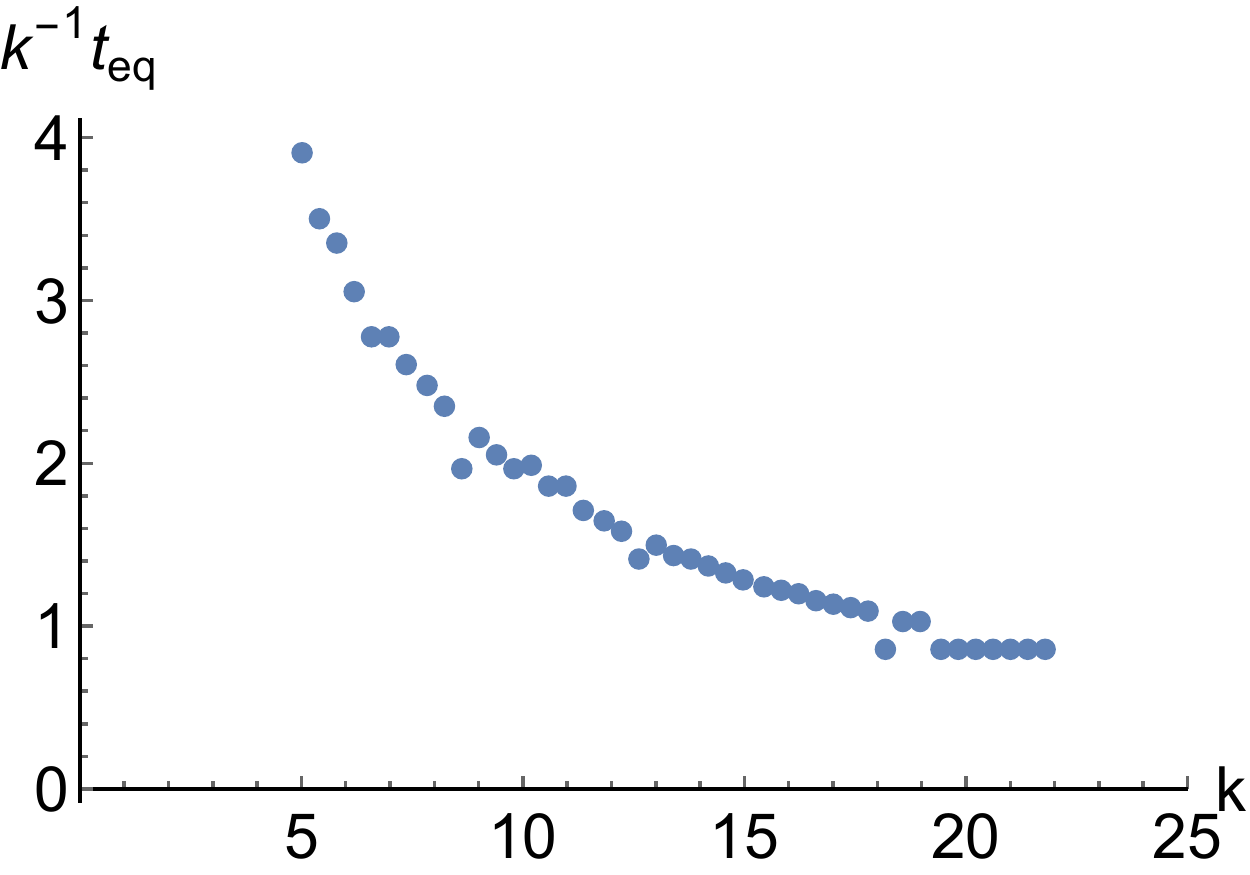}
  \includegraphics[width=65mm]{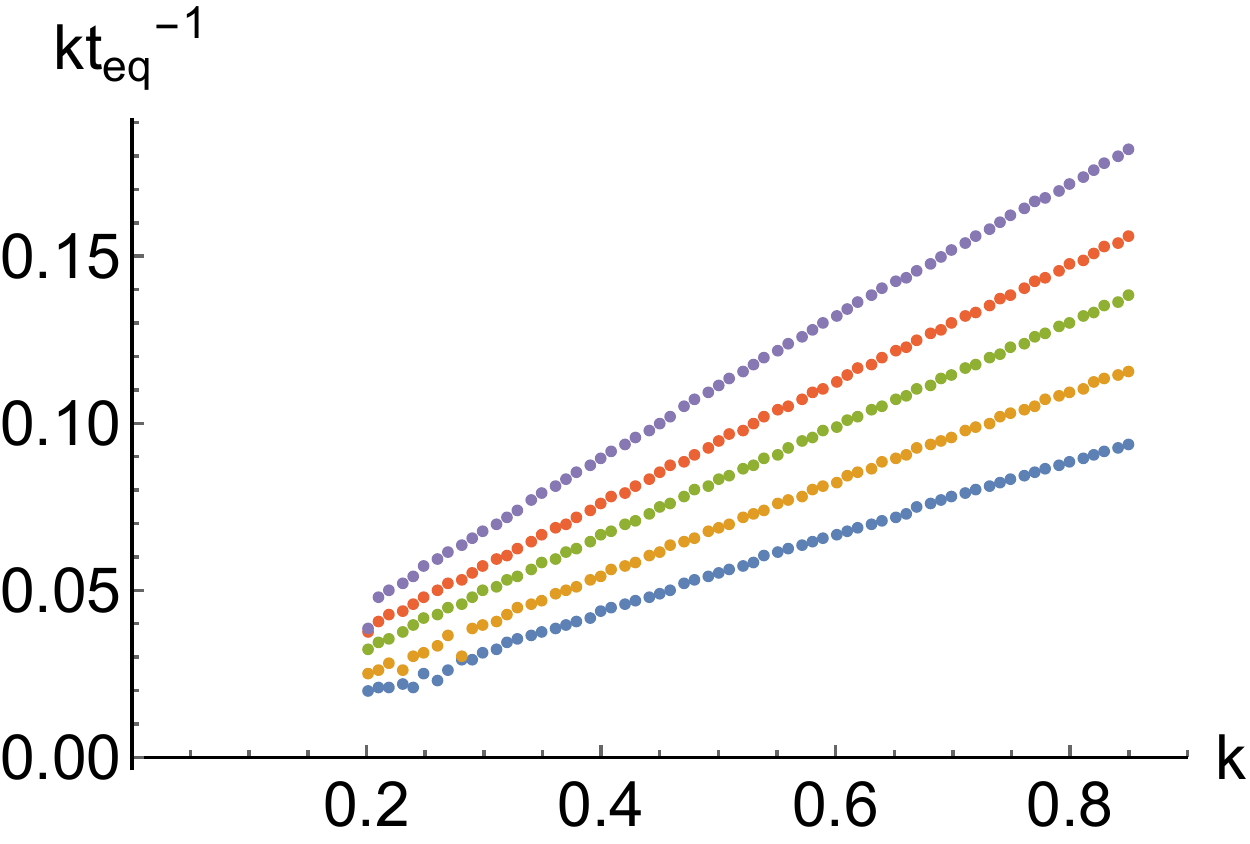}
  \hspace{4mm}
  \includegraphics[width=65mm]{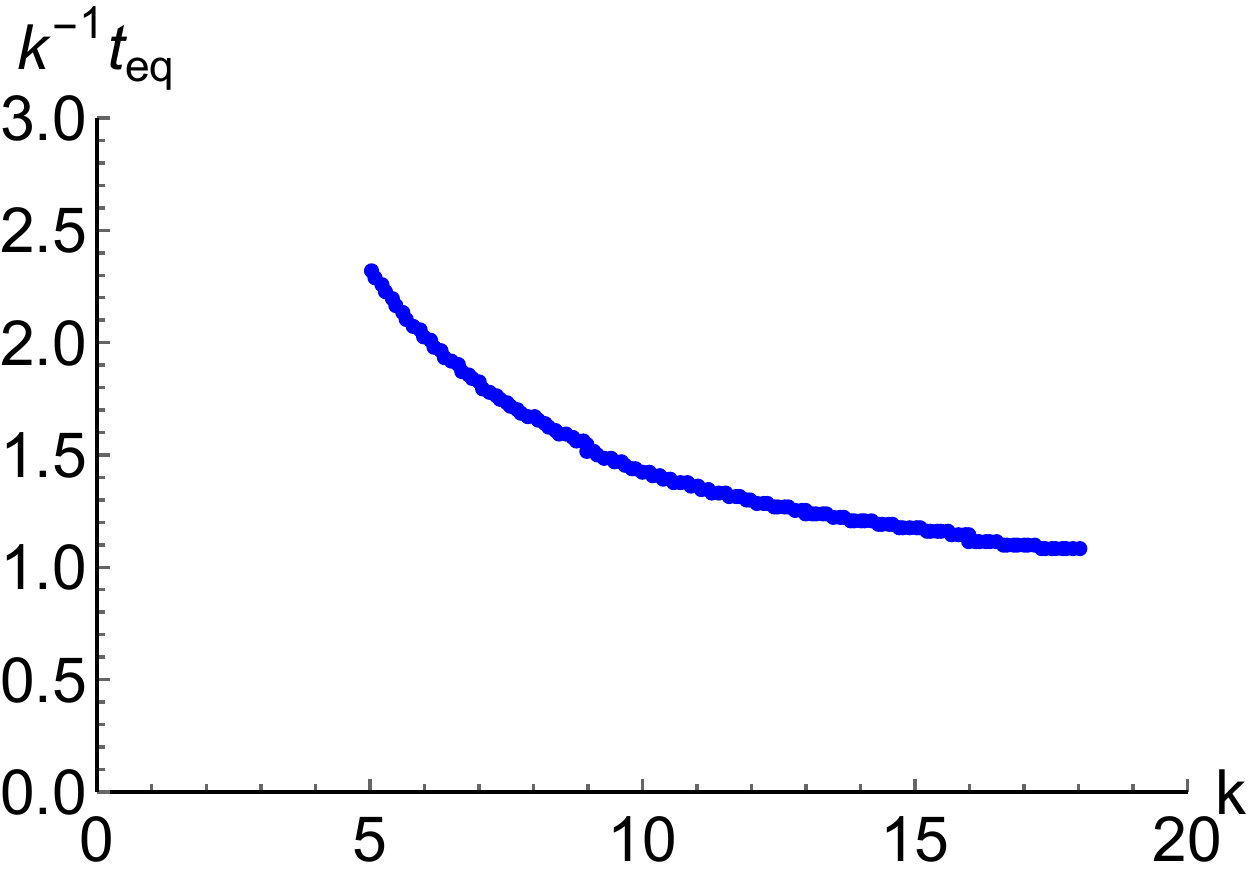}
\caption{Numerical results for the (inverse of) resacled equilibration time, (left, fast quench) right, slow quench,  figure, versus the transition time. For these figures we consider $\epsilon(t_{eq})<0.1$.\\
Top (D2-D4): left plotted for $E_0=0.05, 0.03, 0.02, 0.01$ (top to bottom) and right plotted for $E_0=0.01$. \\
Middle (D3-D5): left plotted for $E_0=0.05$ (orange), 0.03 (green), 0.01 (blue) and  right plotted for $E_0=0.01$. Note that there is a lot of overlaps among different colors. \\
Bottom (D4-D6): left plotted for $E_0=0.05, 0.04, 0.03, 0.02, 0.01$ (top to bottom) and right plotted for $E_0=0.01$. }\label{D4D6fast}
\end{center}
\end{figure}

\begin{itemize}
\item\textbf{D2-D4 System}\\
From \eqref{config}, it is easy to see that the supersymmetric brane configuration for this case is given by 
\be\label{config35}
\begin{array}{cccccccccccc}
& t & x_1 & x_2 & \rho & \Omega_2 & \sigma & \Omega_3  \\
D2 & \times & \times & \times &  &  & &  &  &   &   &  \\
D4 & \times & \times &  &  & \times & \times &  &  &  &  & ,
\end{array}
\ee 
where the SYM degrees of freedom live in $2+1$ dimensional space-time. However, in order to preserve supersymmetry at zero temperature the fundamental matter is confined to $1+1$ dimensions. 
 
\item\textbf{D3-D5 System}\\
The brane configuration \eqref{config} tells us that the supersymmetric brane configuration is 
\be\label{config35}
\begin{array}{cccccccccccc}
& t & x_1 & x_2 & x_3 & \rho & \Omega_2 & \sigma & \Omega_2  \\
D3 & \times & \times & \times & \times &  & &  &  &   &   &  \\
D5 & \times & \times & \times &  & \times & \times &  &  &  &  & ,
\end{array}
\ee 
where the SYM degrees of freedom live in $3+1$ dimensional space-time and the fundamental matter is confined to $2+1$ dimensions. 

\item \textbf{D4-D6 System} \\
In this case the supersymmetric configuration of the branes is given by %
\be\label{config46}
\begin{array}{cccccccccccc}
& t & x_1 & x_2 & x_3 & x_4 & \rho & \Omega_2 & \sigma & \Omega_1  \\
D4 & \times & \times & \times & \times & \times & &  &  &   &   &  \\
D6 & \times & \times & \times & \times &  & \times & \times &  &  &  &.
\end{array}
\ee 
It is clearly seen that although the SYM degrees of freedom live on $4+1$-dimensional space-time, the fundamental matter is confined to $3+1$ dimensions. 
\end{itemize}
We repeat the calculations in the previous section for above cases. In fact the results we are interested inm are the behavior of the (inverse of) rescaled equiliration time in terms of the transition time. These results has been represented in figure \ref{D4D6fast}. They show that, at least for the cases we have considered here, the universal behavior for the fast quenches and adiabatic process for the slow quenches are qualitatively independent of the dimensions in which the SYM or fundamental degrees of freedom live.

%

\section*{Acknowledgment}
M. A. would like to thank School of Particles and Accelerators of Institute for research in fundamental sciences (IPM)  for the research facilities and environment.
The authors would like to thank H. Hashamipour for useful comments.
\appendix
\section{Constant Electric Field in D$p$-D$q$ Scenario}\label{general current}
In order to find $\langle J^x_{eq}\rangle$, as $\langle J^x(t\rightarrow \infty)\rangle= \langle J^x_{eq}\rangle$, one needs to find a solution for \eqref{eom} with
\bse\begin{align}
\label{e1} F_{01}& = \partial_t A_x(t,z) = E_0, \\
\label{e2} F_{0z}& = \partial_z A_x(t,z) = 0, \\
F_{1z}& = \partial_z A_x(t,z) . 
\end{align}\ese %
According to \eqref{e1} and \eqref{e2}, $w$ has no time-dependency and therefore substituting the above equations  in \eqref{eom} yields
\be
 \partial_z\left(\frac{z^{\alpha + \beta}\partial_z A_x}{\sqrt{w}}\right)=0.
\ee
The equilirated current, up to a constant $C$, is defined as \cite{Karch:2007pd}
\be
\langle J^x_{eq}\rangle = C \frac{z^{\alpha + \beta}\partial_z A_x}{\sqrt{w}}.
\ee
Substituting expressions for $\partial_t A_x$ and $\partial_z A_x$ from \eqref{e1} and \eqref{e2} in \eqref{lagranigian} leads to the following
\be
w = \frac{1-\frac{\gamma (2 \pi \alpha ')^2}{R^\alpha}z^\alpha E_0^2}{1-\frac{\gamma (2 \pi \alpha ')^2}{C^2 R^\alpha} z^{-\alpha-2 \beta}\langle J^x_{eq}\rangle^2},
\ee
and by choosing $R=1$ and $2 \pi \alpha ' =1$, one obtains
\be
w = \frac{1-\gamma z^\alpha E_0^2}{1-\frac{\gamma }{C^2} z^{-\alpha-2 \beta}\langle J^x_{eq}\rangle^2}.
\ee
To ensure non-negativity of $w$, we identify the single (real) root of the numerator and the denominator which yields
\be
 \langle J^x_{eq}\rangle = C \gamma ^{- \frac {\beta}{\alpha}}E_0 ^{-1-2 \frac{\beta}{\alpha}}.
\ee
The above formula, of course up to a constant factor, expresses $\langle J_{eq}\rangle$, the equilibrated current, in terms of $E_0$ (which can be considered as the maximum of the imposed time-dependent electric field). For the four cases under study, we present the power dependency of $\langle J^x_{eq}\rangle$ on $E_0$ as follows
\begin{table}[ht]
\caption{$\langle J^x_{eq}\rangle - E_0$ dependency}
\centering
\begin{tabular}{c c}
\hline\hline
$D3 - D7$ & $\langle J^x_{eq}\rangle \propto E_0 ^{3/2}$ \\
$D3 - D5$ & $\langle J^x_{eq}\rangle \propto E_0$  \\
$D4 - D6$ & $\langle J^x_{eq}\rangle \propto E_0^{4/3}$ \\
$D2 - D4$ & $\langle J^x_{eq}\rangle \propto E_0 ^{11/9}$ \\[1ex]
\hline
\end{tabular}
\end{table}

\begin{figure}
\begin{center}
  \includegraphics[width=65mm]{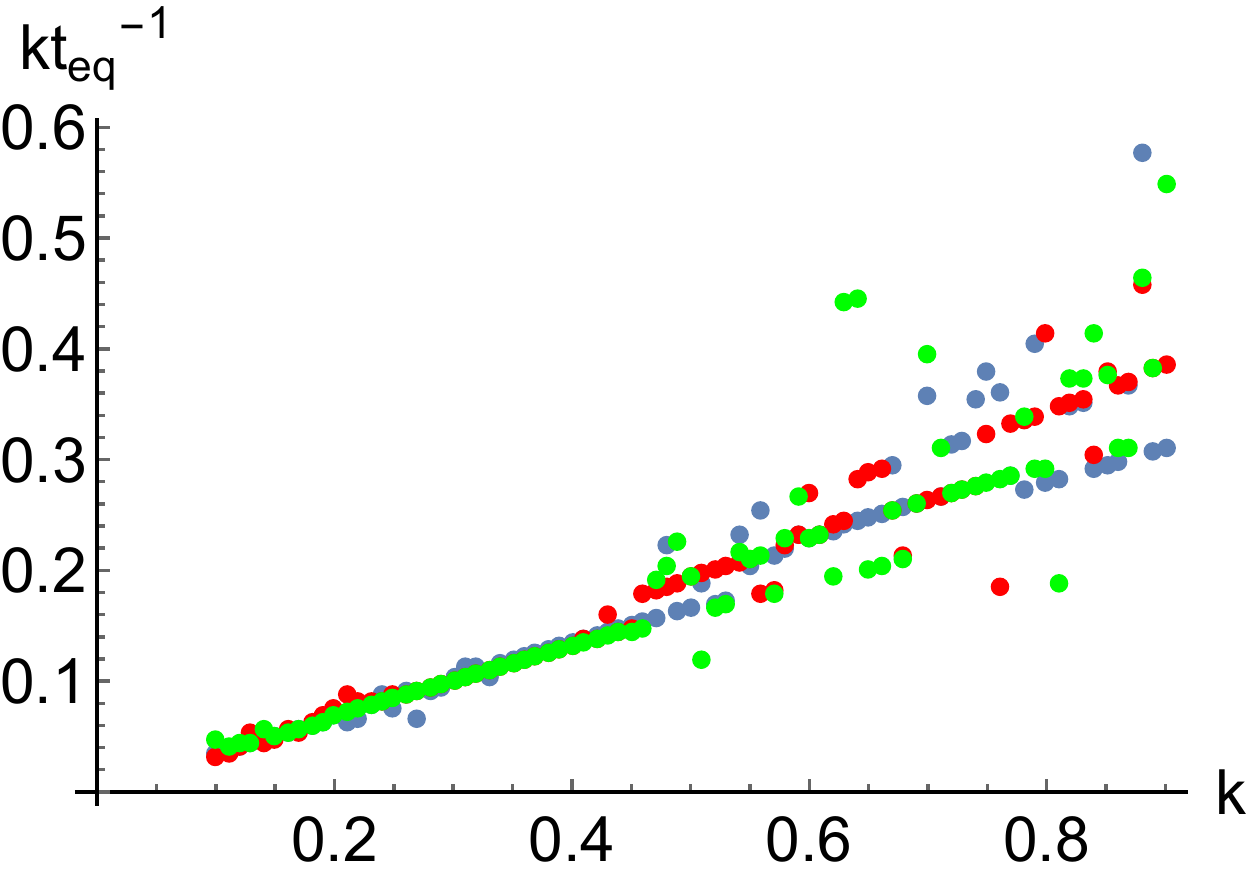}
  \hspace{4mm}
  \includegraphics[width=65mm]{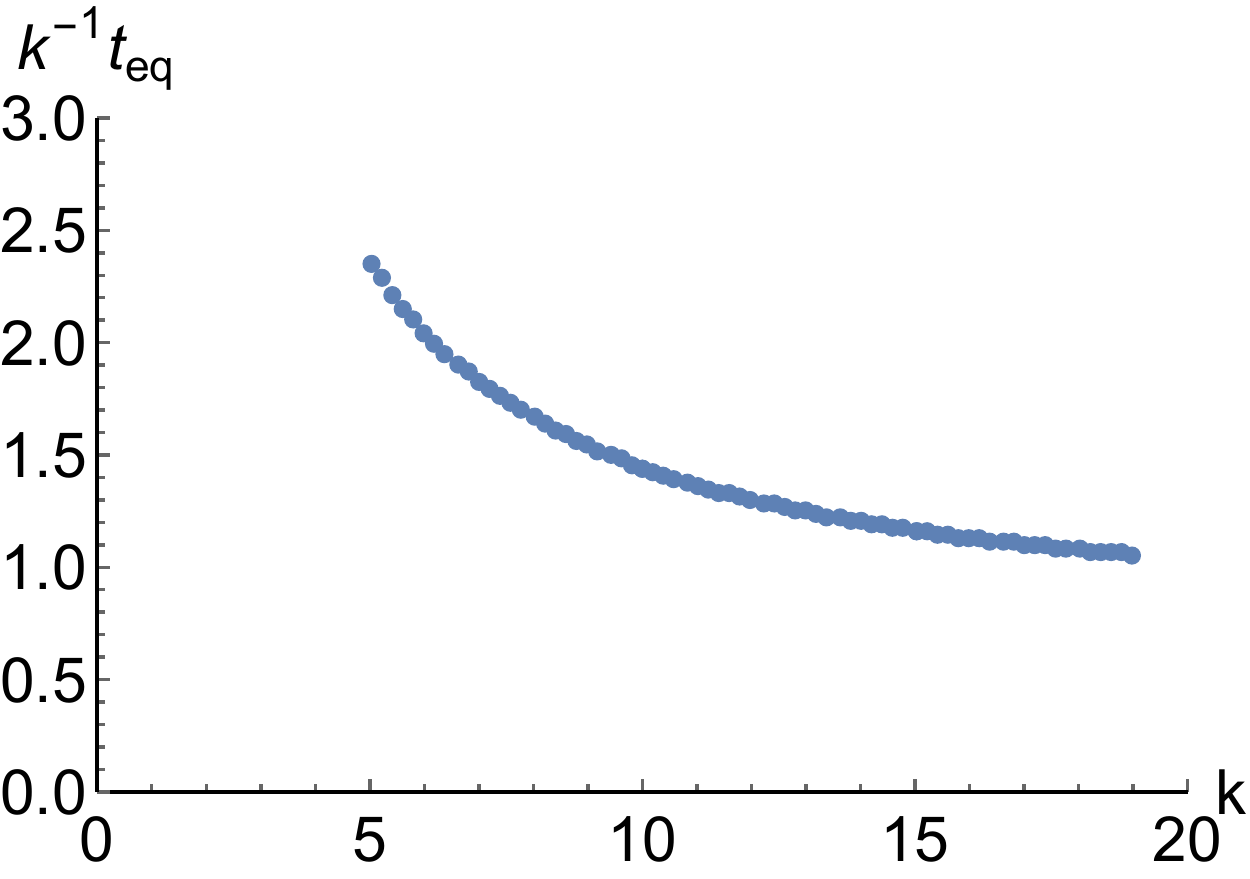}
\caption{Numerical results for the (inverse of) resacled equilibration time, (left) right figure, versus the transition time (for pulse function in \eqref{newpulse}). For these figures we consider $\epsilon(t_{eq})<0.1$.\\
The left panel is plotted for $E_0=0.05, 0.04, 0.03, 0.02, 0.01$ and the right panel is plotted for $E_0=0.01$. }\label{D3D7new}
\end{center}
\end{figure}

\section{Different Pulse Function }
\label{newpulsefunctin}
In this section we would like to investigate whether the behavior of the rescaled equilibration time depends on the pulse function. Thus we will discuss another pulse function which has continuous second derivative. Let us start with %
\be\label{newpulse} %
E(t) =E_0
  \begin{cases}
   0 & \text{if } t \leq 0, \\
   6(\frac{t}{k})^5-15(\frac{t}{k})^4+10(\frac{t}{k})^3 & \text{if } 0 \leq t \leq k ,\\
   1       & \text{if } t \geq k.
  \end{cases}
\ee %
The electric field is zero on $(-\infty,0)$ and starts rising to reach a maximum value at $t=k$. It is easy to repeat the computations in section \ref{37system} and find the numerical results for D3-D7 case. These results are shown in figure \ref{D3D7new}.

\end{document}